\documentstyle[12pt]{article}
\begin{document}

\begin{titlepage}

\begin{center}
{\bf Budker Institute of Nuclear Physics}
\end{center}

\vspace{1cm}

\begin{flushright}
BINP 95-14\\
February 1995
\end{flushright}
\bigskip
\begin{center}
{\bf REGULAR SUPPRESSION\\
OF P,T-VIOLATING NUCLEAR MATRIX ELEMENTS}
\end{center}

\begin{center}
I.B. Khriplovich\footnote{e-mail address: khriplovich@inp.nsk.su}
\end{center}
\begin{center}
Budker Institute of Nuclear Physics, 630090 Novosibirsk,
Russia
\end{center}

\bigskip

\begin{abstract}
In heavy nuclei there is a parametrical suppression, $\;\sim A^{-1/3}\;$,
of T-odd, P-odd matrix elements as compared to T-even, P-odd ones.
\end{abstract}

\vspace{7cm}

\end{titlepage}

Experimental searches for T-odd, P-odd (TOPO) effects in nuclei are
planned now by many groups. To get an idea of their sensitivity one
has to estimate typical value of TOPO mixing matrix elements in
nuclei.

Detailed numerical studies \cite{gv,th} (see also \cite{he}) have
demonstrated that TOPO nuclear matrix elements are regularly smaller
than T-even, P-odd (TEPO) ones. This short note contains a simple
intuitive explanation of this suppression.

We will confine here to a phenomenological treatment of both TEPO and
TOPO interactions. In this approach the effective T-even, P-odd
potential for an external nucleon is presented in a contact form in
the spirit of the Landau-Migdal approach:
\begin{equation}\label{we}
W=\frac{G}{\sqrt{2}}\;\frac{g}{2m}\;\{\vec{\sigma}\vec{p},\rho(r)\}=
\frac{G}{\sqrt{2}}\;\frac{g}{2m}\;\vec{\sigma}[\vec{p}\rho(r)
+\rho(r)\vec{p}\;].
\end{equation}
Here $\{\;, \;\}$ denotes anticommutator, $G=1.027\cdot 10^{-5}
m^{-2}$ is the Fermi weak interaction constant, $m$ is the proton mass,
$\vec{\sigma}$ and $\vec{p}$ are respectively spin and momentum
operators of the valence nucleon, $\rho(r)$ is the density of
nucleons in the core normalized by the condition $\int
d\vec{r}\rho(r)=A$ (the atomic number is assumed to be large, $A\gg
1$). A dimensionless constant $g$ characterizes the strength of the
P-odd nuclear interaction. It is an effective one and includes
already the exchange terms for identical nucleons. This constant
includes also additional suppression factors reflecting long-range
and exchange nature of the P-odd one-meson exchange, as well as the
short-range nucleon-nucleon repulsion. Its typical value is
\begin{equation}
g \sim 1.
\end{equation}

Quite analogously, the effective T-odd, P-odd interaction of an
external nucleon with core can be written as
\begin{equation}
W = \frac{G}{\sqrt 2}\; \frac{\xi}{2m}\; i\, [\vec{\sigma}\vec{p},\rho(r)]
= \frac{G}{\sqrt 2}\; \frac{\xi}{2m}\;\vec{\sigma}\,\vec{\nabla}\rho(r)
\end{equation}
where $[\;,\; ]$ denotes commutator, and $\xi$ is the dimensionless
characteristic of this interaction. The upper limit on the electric
dipole moment of the mercury isotope $^{199}$Hg set in the atomic
experiment \cite{hg}
\begin{equation}\label{hg}
d(^{199}Hg)/e < 9.1 \cdot 10^{-28}\,cm
\end{equation}
bounds this constant as follows:
\begin{equation}
\xi < 1.7\cdot 10^{-3}.
\end{equation}

We are going now to compare nuclear matrix elements of
$\;\{\vec{p},\rho(r)\}\;$ and $\;[\vec{p},\rho(r)]\;$. Let us note first of
all that a heavy nucleus is a semiclassical system, the corresponding
large parameter being $A^{1/3}$. Since the anticommutator has a
classical limit and the commutator does not, it is only natural to
expect that the matrix element of the commutator is suppressed just by
this parameter as compared to that of anticommutator.

Indeed, a simple-minded estimate for the T-even matrix element is
\begin{equation}\label{te}
\langle \{\vec{p},\rho(r)\} \rangle \sim 2 \, \rho_0 \,p_{F}
\sim 3\, \rho_0\, r_0^{-1}.
\end{equation}
Here $\rho_0$ is the nuclear density, $p_{F}$ is the Fermi momentum,
$r_0 = 1.2\;$fm.

Somewhat more intricate is the T-odd case. Assuming for the core
density a step-like profile
\begin{equation}
\rho(r) = \rho_0\,\theta (R-r),
\end{equation}
we get
\begin{equation}
\langle \vec{\nabla} \rho (r) \rangle \sim \rho_0\, \Re^2 R^2
\end{equation}
where $R=r_0\,A^{1/3}$ is the nuclear radius, $\Re$ is the value of
the radial wave function of a valence nucleon at $r = R$. The
following estimate is well-known \cite{bm}
$$\Re^2 R^3 \approx 1.5$$
(it means in fact that the nucleon density at the boundary
constitutes one half of the internal one). In this way we get
\begin{equation}\label{to}
\langle \vec{\nabla} \rho (r) \rangle \sim 1.5 \,\rho_0\, r_0^{-1} A^{-1/3}.
\end{equation}

Of course, the numerical factors in formulae (\ref{te}), (\ref{to})
should not be taken too seriously. However, the parametrical suppression
of the T-odd effect $\;\sim A^{-1/3}\;$ gets obvious.

\bigskip
\bigskip
I am extremely grateful to P. Herczeg for numerous stimulating
discussions, for bringing to my attention Refs. \cite{gv,th}, and for
pushing me to write this note.  Those discussions took place at the
Moriond 95 Workshop; the warm, informal atmosphere of these Workshops
is truly stimulating by itself. This investigation was financially
supported by the Russian Foundation for Basic Research.

\end{document}